\documentclass[prx,amsmath,amssymb,twocolumn,superscriptaddress,showpacs]{revtex4-1}
\usepackage{bm}
\usepackage{amssymb}
\usepackage{colordvi}
\usepackage{graphicx}
\usepackage{color}
\usepackage{hyperref}

\newcommand{\be}{\begin{equation}}
\newcommand{\ee}{\end{equation}}
\newcommand{\bea}{\begin{eqnarray}}
\newcommand{\eea}{\end{eqnarray}}
\newcommand{\ep}{\epsilon}
\newcommand{\vep}{\varepsilon}

\newcommand{\ome}{\omega}
\newcommand{\Ome}{\Omega}

\def\nn{\nonumber}

\def\ket#1{\vert #1 \rangle}

\def\grad {\mbox{\boldmath$\nabla$\unboldmath}}

\begin{document}

\title{3D $Z_2$ Topological Nodes in Nonsymmorphic
  Photonic Crystals: Ultrastrong Coupling and Anomalous Refraction}

\author{Hai-Xiao Wang}\email{These authors contributed equally}
\affiliation{College of Physics, Optoelectronics and Energy, \&
  Collaborative Innovation Center of Suzhou Nano Science and
  Technology, Soochow University, 1 Shizi Street, Suzhou 215006,
  China}
\author{Yige Chen}\email{These authors contributed equally}
\affiliation{Department of Physics, University of Toronto, Toronto,
   M5S 1A7, Canada}
\author{Zhi Hong Hang}
\affiliation{College of Physics, Optoelectronics and Energy, \&
  Collaborative Innovation Center of Suzhou Nano Science and
  Technology, Soochow University, 1 Shizi Street, Suzhou 215006,
  China}
\author{Huanyang Chen}
\affiliation{College of Physics, Optoelectronics and Energy, \&
  Collaborative Innovation Center of Suzhou Nano Science and
  Technology, Soochow University, 1 Shizi Street, Suzhou 215006,
  China}
\author{Hae-Young Kee}\email{hykee@physics.utoronto.ca}
\affiliation{Department of Physics, University of Toronto, Toronto,
   M5S 1A7, Canada}
\affiliation{Canadian Institute for Advanced Research, Toronto, Ontario, M5G 1Z8, Canada}
\author{Jian-Hua Jiang}\email{jianhuajiang@suda.edu.cn}
\affiliation{College of Physics, Optoelectronics and Energy, \&
  Collaborative Innovation Center of Suzhou Nano Science and
  Technology, Soochow University, 1 Shizi Street, Suzhou 215006,
  China}

\date{\today}

\begin{abstract}
We propose to simulate 3D Dirac points and line-nodes with nontrivial
$Z_2$ topology in nonsymmorphic all-dielectric photonic-crystals
with space-time reversal symmetry, which can be realized at infrared
and microwave frequencies. Double degeneracy of all Bloch states in
high symmetry planes is achieved via nonsymmorphic screw symmetries
despite the fundamental obstacle of no Kramers degeneracy 
in photonic crystals. Two orthogonal screw axes lead to 3D $Z_2$ Dirac 
points on high symmetry Brillouin zone boundary lines. On the other
hand, $Z_2$ line-nodes emerge as protected twofold degeneracy 
of Bloch bands with opposite mirror parities on the $k_z=0$ 
plane. The lowest frequency line-node is deterministic because of
a degenerate-partner switching mechanism guaranteed by the fundamental
properties of Maxwell equations and the nonsymmorphic screw symmetry. 
A pair of Fermi arcs with opposite chirality due to $Z_2$ topological
Dirac points emerge below the light-line on (100) and (010)
photonic-crystal-air interfaces. These robust surface states offer an 
unique opportunity to realize an ``open cavity'' with strong
interaction between quantum emitters and engineered vacuum with
nontrivial Berry phases --- an important step toward 
topological states of strongly interacting bosons. Realistic
calculation for resonant coupling between cavity-photons and 
phonons in boron nitride thin film yields ultrastrong coupling with 
vacuum Rabi splitting reaching to $23\%$ of photon frequency.
We also show that type-II Dirac cones have anomalous valley selective
refraction: birefringence with both positive and negative refractions
for one valley, while no refraction for the other.
\end{abstract}


\maketitle

\section{Introduction}
The study of topological properties of energy bands lies at the 
heart of frontiers of condensed matter physics, ranging from
topological insulators\cite{ti1,ti2}, topological
superconductors\cite{ti2}, topological semimetals\cite{tsm,tsm2,tsm3} and
symmetry-protected topological states in strongly correlated quantum
materials\cite{wenbook,wenscience}. Recent discoveries of
Dirac\cite{wang,dirac,mob} and Weyl semimetals\cite{ling-exp,weyl1,weyl2}
with ultrahigh mobility and other exotic properties 
bring those developments closer to material science and
applications. 

In parallel, photonics is another realm where achieving robust
transport\cite{john,haldane1,mit1,taylor},
selective waveguiding\cite{nnano}, and manipulation of angular
momentum\cite{nphoton} are of central importance. These are the major
challenges in photonic circuits, metamaterials, and quantum
optics. Photonic topological states introduce new routes to overcome
those challenges. For instance, back-scattering immune edge states of
photonic quantum Hall systems are perfect solutions for robust and
unidirectional waveguiding\cite{mit2,kejie,hafezi1,ling1,floquet}.
Recent developments on manipulation of light propagation via
polarization and angular momentum request strong spin-orbit coupling
and angular-momentum--wavevector locking\cite{nphoton}. These
properties are available from topological nodes in photonic
bands\cite{ct-2d,z2p,ctti,huxiao,lumh,our,szhang}. Right 
at the heart of photonic topological states, there are
nontrivial Berry phases and robust surface states, which offer
unprecedented control of light and novel
physical effects\cite{mit2,ling-exp,kejie,floquet,ct-zak,ct-exp,hafezi2,hafezi3,lingrev}.

In this work, we propose to simulate 3D $Z_2$ topological Dirac points
(DPs) and line-nodes in
nonsymmorphic {\em all-dielectric} photonic crystals (PhCs) with
space-time reversal symmetry. All-dielectric PhCs can be fabricated with
well-controlled geometries from microwave to near-infrared 
frequencies. Disorder and interaction effects that complicate
electronic band structures can be very weak in all-dielectric PhCs.
Compared with plasmonic metamaterials or
magneto-optical materials\cite{z2p,lumh}, all-dielectric PhCs are
lossless materials that preserve photonic coherence at
large-scales and long-times. Future developments of topological
photonics will benefit from simple all-dielectric PhCs, where
photonic $Z_2$ topology becomes very important\cite{huxiao,our}.

However, the fundamental difference between photon and electron, is
that photon, as a boson, does not have Kramers degeneracy.
It is thus rather challenging to create $Z_2$ topological states in
all-dielectric PhCs\cite{huxiao,our}. Up till now, there are very few
studies on 3D $Z_2$ topological states in PhCs, based on magneto-optical
materials with glide symmetry\cite{ling2} or point group
symmetry\cite{our}. In this work, we show that photonic $Z_2$
topological nodes can be created using nonsymmorphic screw
symmetries. Specifically, screw symmetries generate twofold degeneracy
for all Bloch states on certain high symmetry planes of the Brillouin zone
(BZ). This mechanism, similar to Kramers degeneracy 
in fermionic systems, leads to rich phenomena in band degeneracy. As a
consequence, a number of 3D $Z_2$ DPs and $Z_2$ line-nodes are created,
unveiling effective methodology toward photonic $Z_2$ topological states.

We found the ``Fermi arcs'' associated with $Z_2$
DPs, revealing that the surface states are a pair of mirror
symmetric, helically dispersed bands with opposite chirality. These
robust double-Fermi-arc surface states are readily observable
in microwave experiments. We further show that part of the surface
states survive on the (100) and (010) PhC-air interfaces, since they emerge
below the light-line. Similar confinement of surface states associated
with $Z_2$ line-nodes appear on the (001) PhC-air interface. Such
surface states offer access to ``open cavity'' photonic states with
nontrivial Berry phases. Placing hexagonal 
boron-nitride multilayers brings in {\em ultrastrong} light-matter
interaction with vacuum Rabi splitting reaching to $23\%$ of the
optical phonon frequency, as induced by significant enhancement
of electromagnetic fields at the interface. 
This system offers a practical route toward ultrastrong coupling
regime for mid infrared frequency where the quantum nature of photon becomes
significant\cite{ciutirmp,ghz,thz1,thz2,mir,qve}. In addition, this 
scheme opens the possibility toward simulation of quantum bosonic
systems with both nontrivial Berry phases and strong
interaction, where topological states of strongly interacting photons may
emerge\cite{wenscience,fqhem}. Moreover, we find that type-II $Z_2$ 
DPs have anomalous valley-selective refraction, providing a
path toward valley physics for Dirac photons.

\section{Nonsymmorphic all-dielectric Photonic crystals}
To demonstrate creation of 3D $Z_2$ topological nodes using nonsymmorphic
symmetries, we study a simple, all-dielectric PhC of which the
structure is illustrated in Figs.~1(a), 1(b), and 1(c). It is 
a tetragonal lattice with space group of P4$_2$/mcm.
In each unit cell (with lattice constant
$a$ along all three directions) 
there are two orthogonal dielectric blocks (painted as yellow and
green in the figures) of the same shape that are embedded in the background
medium. Each of them has length $l=0.5a$, width $w=0.2a$, and height
$h=0.5a$, separately. The centers of the blocks are at $(0, 0, 0)$ and
$(0.5a, 0.5a, 0.5a)$, respectively (origin of the 
coordinate is defined as the center of the unit cell). The blocks have
(relative) permittivity $\vep_s=24$, while the background medium has
permittivity $\vep_b=2.2$. The blocks can be made of high-index
materials such as tellurium (for wavelength longer than 4~$\mu$m),
whereas the background material can be polymers such as PMMA.

\begin{figure}
\begin{center}
\includegraphics[width=8.6cm]{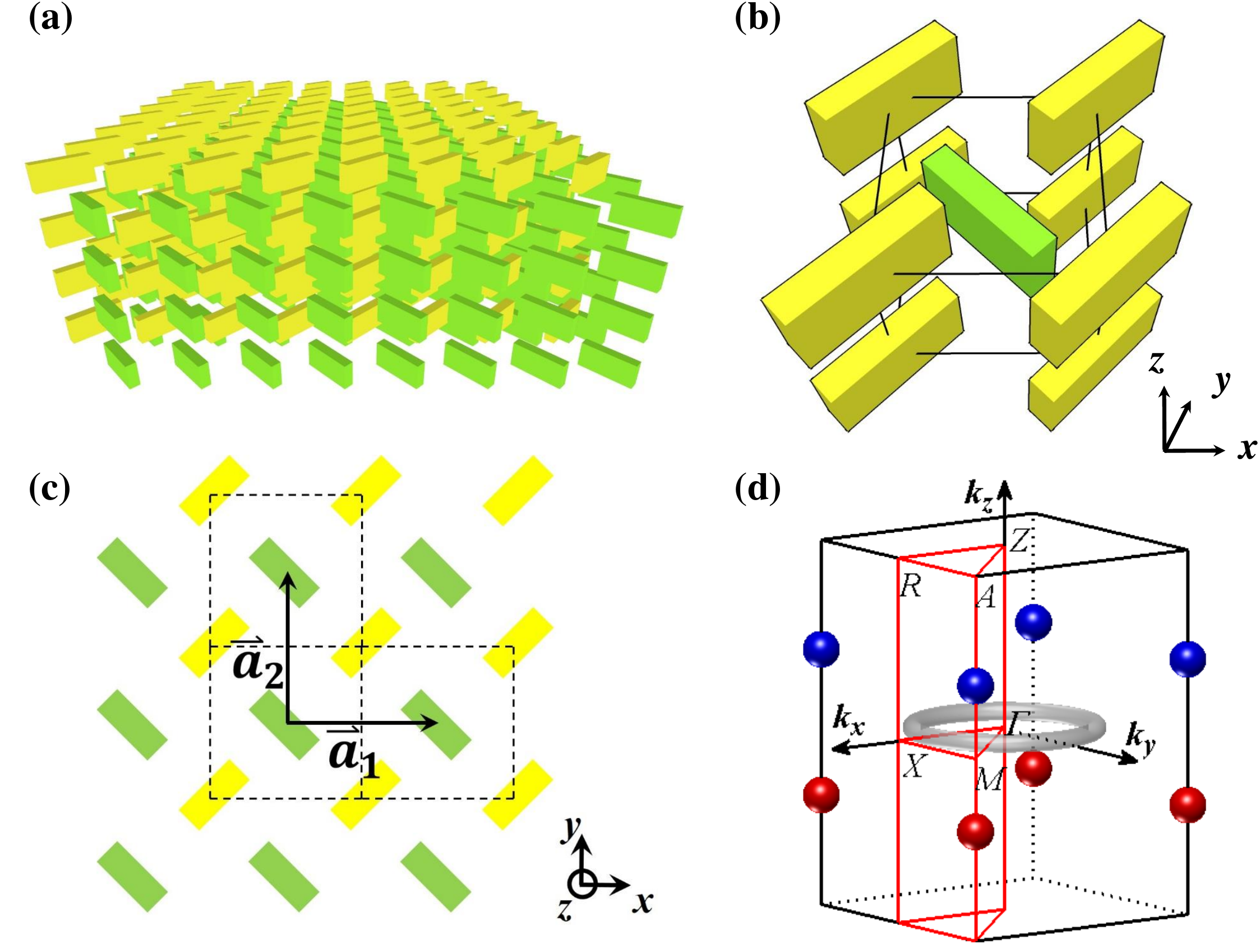}
\caption{ (Color online) (a) Tetragonal lattice photonic crystal with
  nonsymmorphic symmetry. (b) Structure of a real space unit
  cell. The unit cell boundary is illustrated by the black lines. (c) Top-down
  view of the photonic crystal. Yellow and green 
  blocks are of the same material and shape. $\vec{a}_1$ and $\vec{a}_2$ are the
  lattice vectors in the $x$-$y$
  plane. $|\vec{a}_1|=|\vec{a}_2|=a$. (d) First BZ with a pair of 
  DPs on the M-A line and a line-node around the $\Gamma$
  point in the $k_z=0$ plane. Red and blue DPs have opposite
  $Z_2$ topological charge.}
\end{center}
\end{figure}

The tetragonal lattice PhC possesses the following elemental symmetries,
\bea
({\rm inversion})\ {\cal P} &:& (x,y,z)\rightarrow (-x,-y,-z) ,\nn\\
({\rm glide})\ G_x &:& (x,y,z)\rightarrow (0.5a-x, 0.5a+y, 0.5a+z)
,\nn\\
({\rm mirror})\ M_1 &:& (x,y,z)\rightarrow(y,x,z), \nn\\
({\rm mirror})\ M_2 &:& (x,y,z)\rightarrow (-y,-x,z) .
\eea
and their combinations. For instance, there are another glide
symmetry and two screw symmetries,
\bea
G_y = G_x C_{2z}, \quad S_x = G_x {\cal P}, \quad S_y = G_y {\cal P}
. \label{screw}
\eea
It is also important to define the 180$^\circ$ rotation along $z$ axis
\be
C_{2z}\equiv M_1M_2 : (x,y,z) \rightarrow (-x,-y,z) ,
\ee
and the mirror symmetry with respect to the $z=0$ plane,
\be
M_z \equiv {\cal P} C_{2z} : (x, y, z) \rightarrow (x, y, -z) .
\ee
The nonsymmorphic symmetries transform the yellow blocks into green
blocks (and vice versa), while the symmorphic symmetries transform
within each type of blocks. In PhCs, since for the same band the Bloch
functions of the electric field and the magnetic field carry the same
symmetry information, we will henceforth use only the magnetic fields
${\vec h}$ for the discussions on the symmetries of photonic
bands. The magnetic field is a pseudovector which 
is even under space inversion but odd under time-reversal
operation. Therefore, the explicit form of time-reversal operator for
the magnetic field is ${\cal T}=-{\cal K}$ where ${\cal K}$ is the complex
conjugation operator. Here we shall shortly denote the
time-reversal operator as
\be
{\cal T} : t \to -t .
\ee
The apparent parities (i.e., mirror symmetries) of the magnetic fields
are also useful for constructing the ${\vec k}\cdot{\vec P}$ theory of
photonic bands around the topological nodes [see
Appendix~A]. Numerical calculation in this work is largely based on
the open software ``MIT Photonic Bands'' (see Ref.~\cite{mpb}). For
tetragonal lattice, the spectrum and other properties remain the same 
if $k_x$ and $k_y$ are interchanged. We thus only discuss one of them
in this work.

Since both time-reversal and inversion symmetry exist, the $U(1)$
Berry phase vanishes. There can only be $SU(2)$ Berry phases and $Z_2$ 
topological states. For topological nodes, there can only be $Z_2$
DPs and line-nodes, whereas Weyl points cannot
emerge in our PhC (unless inversion symmetry is broken).

\section{Dirac Points and Fermi arc surface states}

\begin{center}
  \begin{figure}
    \includegraphics[width=8.6cm]{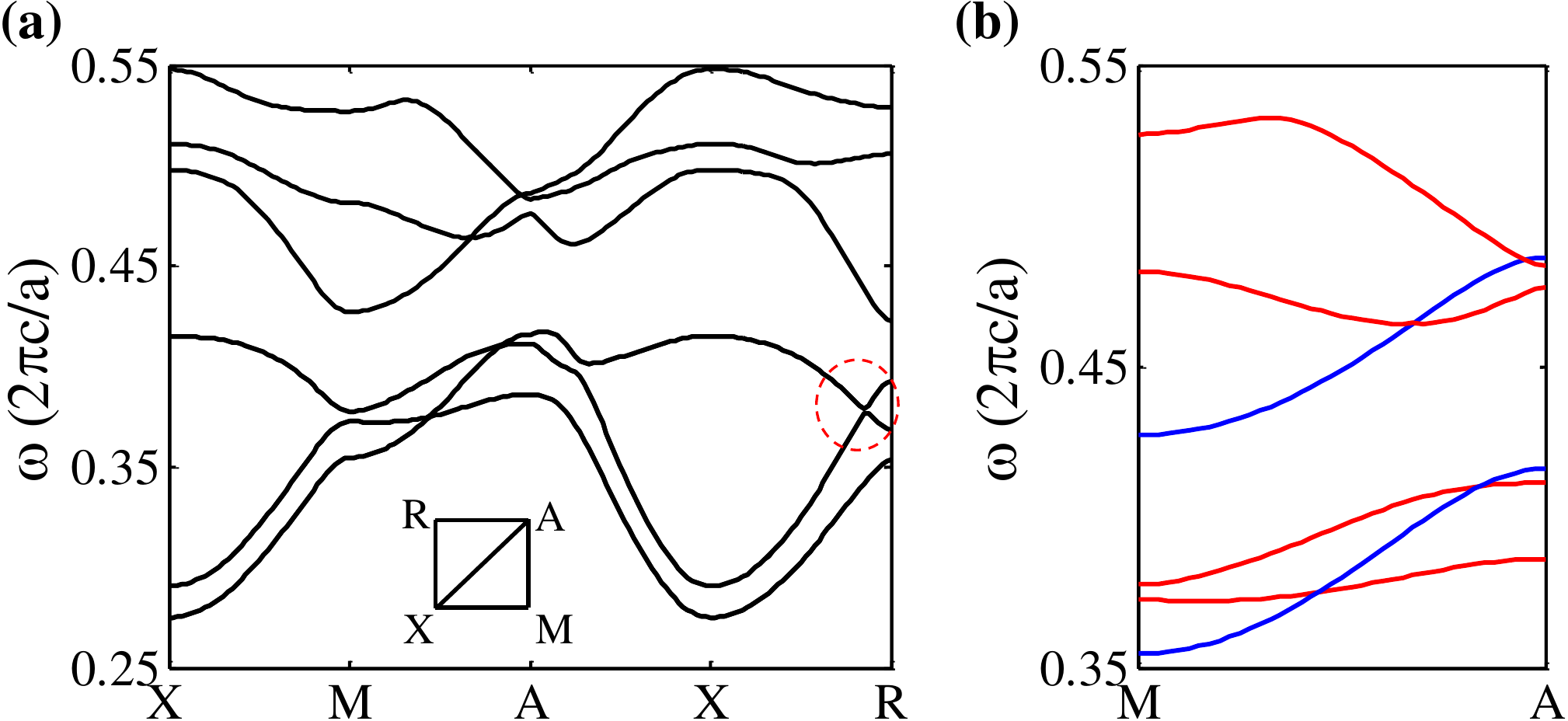}
    \caption{ (Color online) (a) Band structure for the $k_x=\frac{\pi}{a}$
      plane. The dashed red circle contains an anti-crossing of two
      sets of doubly degenerate bands. (b) Band dispersion along the M-A
      line. In this line, the photonic bands are eigen functions of the $C_{2z}$ operator. The
      eigenvalues can be -1 (odd parity) or 1 (even parity), as
      labeled by red and blue in (b), respectively. Crossing of bands
      with opposite parity yields DPs. Frequencies are in unit of $2\pi c/a$
      with $c$ being the speed of light in vacuum.}
  \end{figure}
\end{center}

\subsection{Band degeneracy from screw symmetries}
We now show that nonsymmorphic screw symmetries are effective tools toward
doubly degenerate photonic bands. In our PhCs, the twofold degenerate
bands are induced by the following {\em anti-unitary} symmetry operators: 
\begin{align}
& \Theta_x\equiv S_x {\cal T} : (x,y,z,t)\rightarrow \nn\\
& \quad \quad \quad \quad \quad \quad (0.5a+x, 0.5a-y, 0.5a-z, -t) , \nn\\
& \Theta_y\equiv S_y {\cal T} : (x,y,z,t)\rightarrow \nn\\
& \quad \quad \quad \quad \quad \quad (0.5a-x, 0.5a+y, 0.5a-z, -t) , 
\end{align}
where $S_x$ and $S_y$ are the screw operations defined in
Eq.~(\ref{screw}). It is straightforward to show that for all Bloch states,
of the form $\Psi_{n{\vec k}}({\vec r})=e^{i{\vec k}\cdot{\vec
    r}}u_{n{\vec k}}({\vec r})$ ($u$, can be a vector, is a periodic
function of lattice translation, while $n$ is the band index), 
\be
{\Theta}_x^2 \Psi_{n{\vec k}}({\vec r}) =e^{-i k_xa} \Psi_{n{\vec k}}({\vec r}).
\ee 
The $\Theta_x$ operator transform $(k_x, k_y, k_z)$ to $(-k_x, k_y,
k_z)$. On the $k_x=\frac{\pi}{a}$ plane, which is
invariant under $\Theta_x$, one has
\be
{\Theta}_x^2=-1 .
\ee
Therefore, all eigenstates on the $k_x=\pi/a$ plane is twofold
degenerate, following similar arguments of the Kramers degeneracy in
fermionic systems. Similarly, for the $k_y=\frac{\pi}{a}$ plane, {\em 
  all} Bloch states are doubly degenerate due to $\Theta_y^2=-1$. In
tetragonal lattice, the spectrum is same for the
$k_y=\frac{\pi}{a}$ plane as for the $k_x=\frac{\pi}{a}$ plane [shown
in Fig.~2(a)]. The above results demonstrate the importance of
nonsymmorphic screw symmetry in simulation of photonic $Z_2$
topological states.

\begin{figure}
\begin{center}
\includegraphics[width=8.6cm]{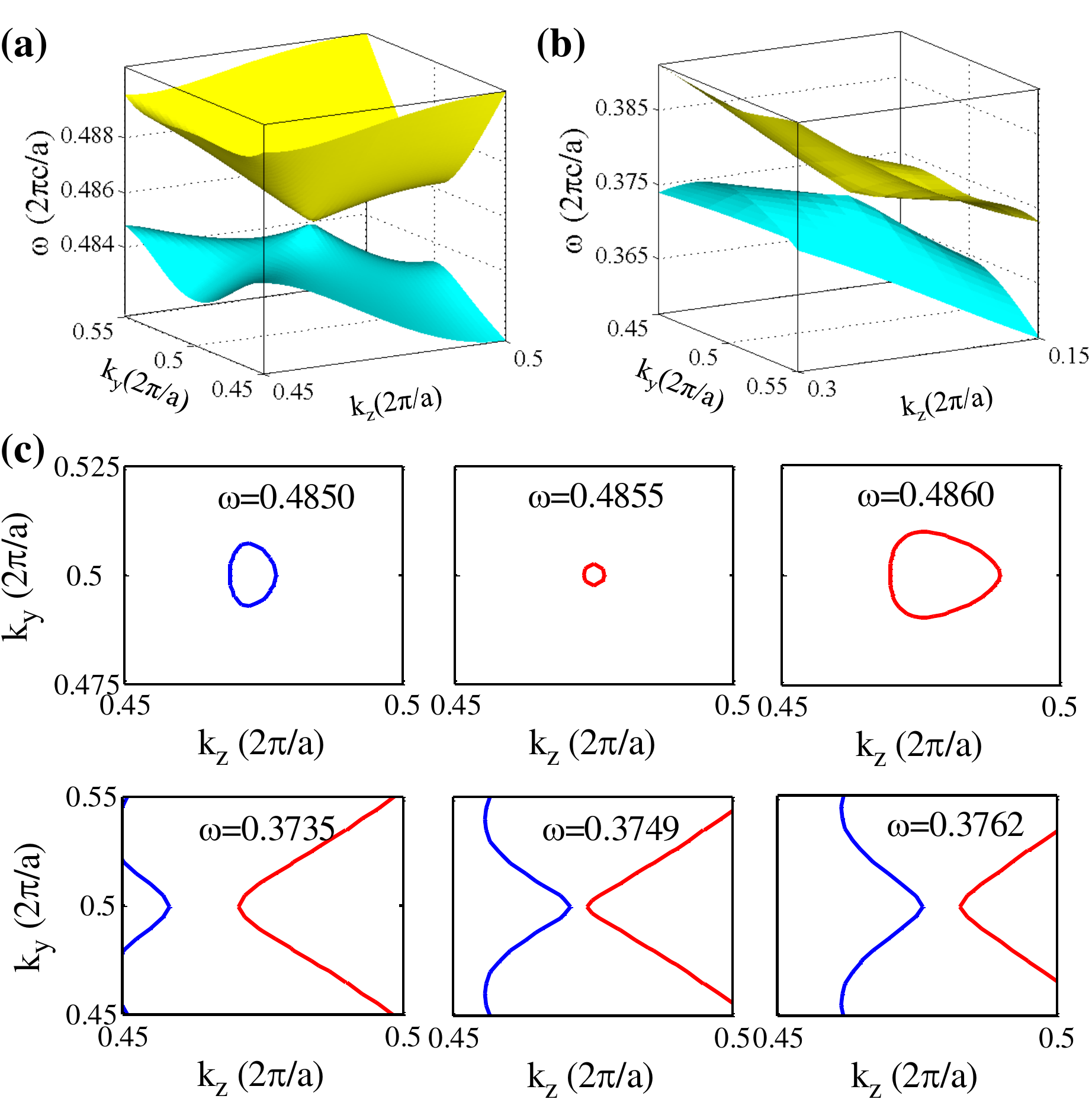}
\caption{ (Color online) Photonic dispersion around a (a) type-I DP and
  (b) type-II DP in the $k_y$-$k_z$ plane. The type-I DP is the fourth
  DP, while the type-II DP is 
  the first DP, counting from low to high
  frequency. Isofrequency contours of the (c) type-I and (d) type-II Dirac cones.}
\end{center}
\end{figure}

\subsection{Dirac Points on M-A line}
DPs are fourfold degenerate points with linear-dispersion
where the effective Hamiltonian describing the photonic states
resembles that of massless Dirac equation [The effective Hamiltonian
can be obtained via ${\vec k}\cdot{\vec P}$ expansion of the Maxwell
equation around the DP (see Appendix~A)]. A DP has
vanishing Chern number but may carry nontrivial $Z_2$ topology.
The topological charge of a DP can be determined by
the eigenvalue of the rotation symmetry that protects the DP as shown
in Ref.~\cite{yang}. $Z_2$ topological DPs 
appear in pairs (at two ${\vec k}$ points linked by the time-reversal
operation) with {\em opposite} $Z_2$ topological charge.

The two planes $k_x=\frac{\pi}{a}$ and $k_y=\frac{\pi}{a}$ share a
common line, i.e., the M-A line:
$(\frac{\pi}{a},\frac{\pi}{a},k_z),~\forall k_z$. 
It is crucial to notice that
\be
\Theta_y \Theta_x = e^{-i(k_x+k_y)a} \Theta_x \Theta_y . \label{thxy}
\ee
Therefore, $\Theta_x$ commutes with $\Theta_y$ on the M-A
line. It makes sense that $\Theta_x\Psi_{n{\vec k}}({\vec r})$ is the
same state as $\Theta_y\Psi_{n{\vec k}}({\vec r})$ on this line, since
the Bloch states here are invariant under both $M_1$ and $M_2$ transformations.
Moreover, on the M-A line $C_{2z}$ is a symmetry operation, and
\be
\Theta_x \Theta_y = \left .e^{-ik_xa} C_{2z}\right|_{k_xa=\pi} = - C_{2z}.
\ee
Thus $C_{2z}$ commutes with both $\Theta_x$ and
$\Theta_y$. As a consequence, each doublet (i.e., $\Psi_{n{\vec
    k}}({\vec r})$ and $\Theta_x\Psi_{n{\vec k}}({\vec r})$) has
the {\em same} $C_{2z}$ eigenvalue (denoted as $c_{2z}$). Since
$C_{2z}$ is 180$^\circ$ rotation around $z$ axis, its eigenvalue
$c_{2z}=\pm 1$ represents parity in the $x$-$y$ plane.
Crossing between two sets of doublets with opposite parities yields a
DP protected by the $C_{2z}$ symmetry. The $Z_2$ topological
charge measures parity inversion across the DP.
For a DP at $(\frac{\pi}{a},\frac{\pi}{a}, k_{DP})$, the
$Z_2$ topological charge is given by\cite{yang}
\be
{\cal N}_{DP} = \frac{1}{2}[c_{2z}^{-}(k_{DP}^+) - c_{2z}^{-}((k_{DP}^-)]
\ee
where $k_{DP}^+=k_{DP}+0^+$ and $k_{DP}^-=k_{DP}-0^+$ are the
wavevectors slightly larger or smaller than $k_{DP}$,
respectively. Here $c_{2z}^{-}$ denotes the $C_{2z}$ eigenvalue of
the lower doublet near the DP. Therefore, a DP is a
source or drain of parity inversion. For our PhCs, there are several
DPs along the M-A line. The parity of those bands are
labeled by colors in Fig.~2(b). We find that the lowest four DPs with $k_z>0$ all carry topological charge ${\cal N}_{DP}=-1$,
while the lowest four DPs with $k_z<0$ all carry topological
charge ${\cal N}_{DP}=1$. 

We remark that the above scenario does not apply for the X-R line
($k_x=\pi/a$ and $k_y=0$). On these lines, according to Eq.~(\ref{thxy}),
$\Theta_x$ anti-commutes with $\Theta_y$ since $k_x+k_y=\pi/a$.
The degenerate partners in each doublet then have
opposite $c_{2z}$. Hence the two sets of doubly degenerate bands
can anti-cross each other, as observed in our calculation.

\begin{figure}
\begin{center}
\includegraphics[width=8.6cm]{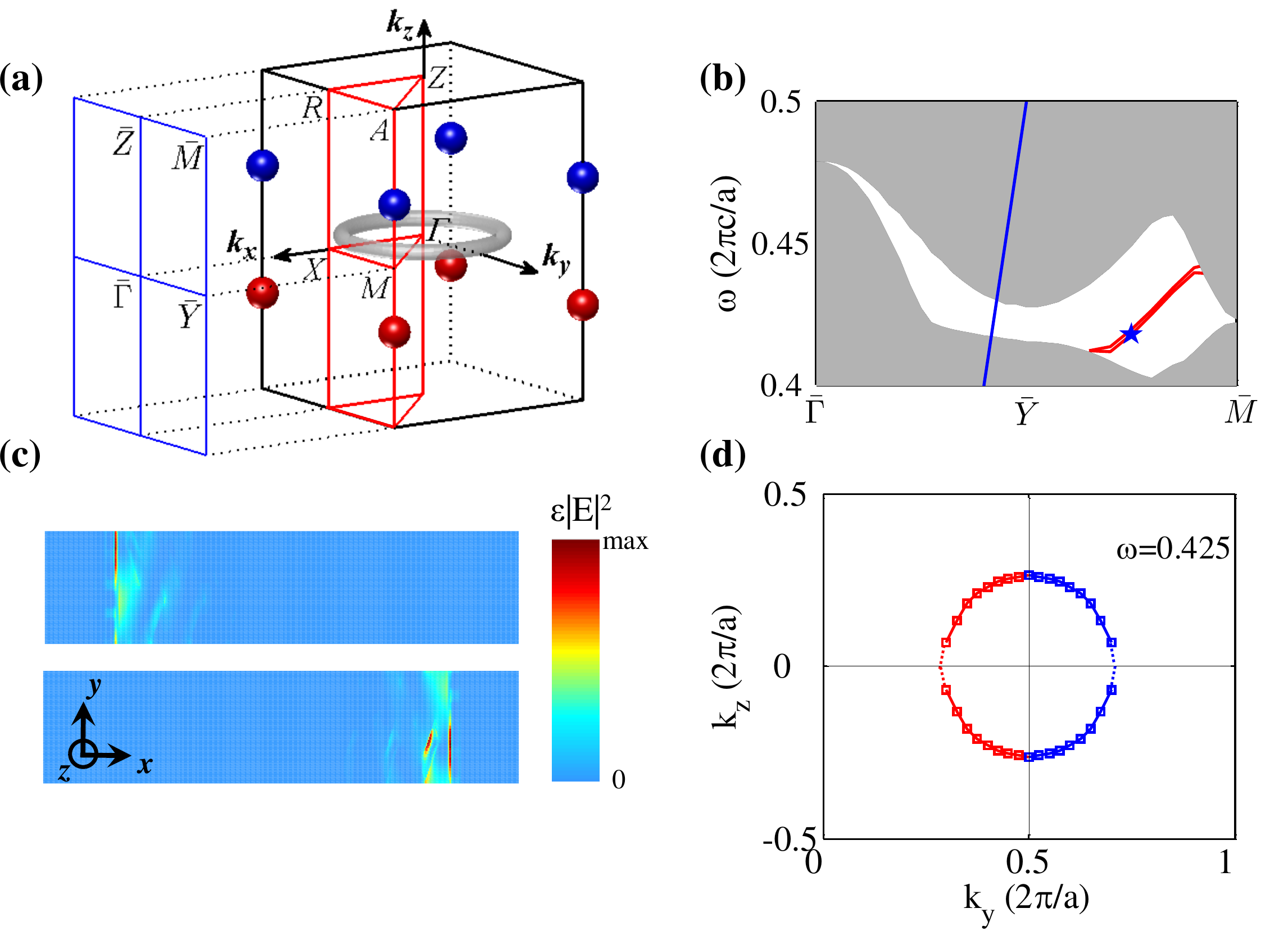}
\caption{ (Color online) (a) DPs and surface Brillouin
  zone for (100) surface. Red DPs have topological charge ${\cal
    N}_{DP}=1$, while blue ones have ${\cal N}_{DP}=-1$. The
  supercell is stacked along $x$ direction, so that the surface BZ is
  projected to the $k_y$-$k_z$ plane. (b) Dispersion of topological
  surface states (red) as well as the projected bulk photonic bands
  (gray). Two nearly degenerate states are labeled by a blue
  star. Light-line is represented by a blue line.
  (c) Electromagnetic energy profiles for the two surface
  states in the $x$-$y$ plane. The wavevector of the edge states is
  $(0, 0.5, 0.25)\frac{2\pi}{a}$. (d) Double ``Fermi arcs'' due to
  $Z_2$ DPs for $\ome=0.425(\frac{2\pi c}{a})$. The two Fermi arcs
  (labeled as red and blue)
  intersect with each other when $k_y=\pi/a$. }
\end{center}
\end{figure}

We notice from Fig.~2 that there are both type-I and type-II Dirac
cones. To explore the properties of these DPs, we plot their
dispersion on the $k_y$-$k_z$ plane in Figs.~3(a) and 3(b). Fig.~3(a)
shows a Dirac cone with slightly tilted dispersion (i.e., type-I),
while Fig.~3(b) shows a highly tilted Dirac cone (i.e., type-II). The
difference between the two Dirac cones is further displayed in
Figs.~3(c) and 3(d). Here Fig.~3(c) shows that a type-I DP
has closed isofrequency contours. Approaching the DP
the circumference of the isofrequency contour goes to zero. In
contrast, a type-II Dirac cone, as shown in Fig.~3(d), exhibits 
isofrequency contours consisting of two unconnected branches. In
approaching the frequency of the DP, the two branches touch
each other at the DP. Therefore, the density of states can be
much greater for type-II DPs which may lead to enhanced
light-matter interaction. We shall show in Sec.~\ref{sec:app}~B that
type-II DPs have anomalous valley-selective refraction as well.

\subsection{Double Fermi arc surface states on (100) surface}

According to the projection principle of topological
nodes\cite{pra,ct-exp,wang,cfang,tsm,tsm2,tsm3}, we find that the
(100) and (010) surface states are associated with the $Z_2$ DPs. 
According to the symmetry of our PhC, the spectral properties
of those surface states are the same. We hence focus only on the (100)
surface in this section.

We calculate the edge states on the (100) surface of our photonic
crystal using a supercell numerical scheme (for details, see
Appendix~B). The screw symmetry $S_y$, which is essential for double
degeneracy at $k_y=\frac{\pi}{a}$, is kept for the supercell. The
surface BZ is shown in Fig.~4(a). Two almost degenerate surface states
can be identified in Fig.~4(b). Their field profiles in Fig.~4(c)
reveal that those two states are edge states at opposite boundaries. To 
make sure that those are the topological edge states, we calculate the 
``Fermi arcs'' through a scan of the edge states in the surface
BZ (see Appendix~B). The ``Fermi arcs'' for $\ome=0.425\frac{2\pi
  c}{a}$ are plotted in Fig.~4(d). From the figure there are two Fermi
arcs with opposite chirality (labeled as red and blue), which is
significantly different from chiral Fermi arcs in Weyl
PhCs\cite{ct-exp,szhang}. Those two
Fermi arcs intersect with each other when $k_y=\frac{\pi}{a}$.
The degeneracy of the two Fermi arcs on the $k_y=\frac{\pi}{a}$ line
is protected by the $\Theta_y$ symmetry. Moreover, the spectra of the
two Fermi arcs transform into each other under $\Theta_y$ which is
manifested in the surface BZ as $(k_y,k_z)\rightarrow (-k_y,k_z)$.
Such mirror symmetric Fermi arcs is a smoking-gun signature of the
topological surface states associated with $Z_2$ DPs. The surface
states are also found to be quite robust to surface truncation geometries
(see Appendix~C), demonstrating topological protection. The surface
states shown here are similar to the double helicoid surface states
predicted recently in an electronic material with different
symmetry\cite{cfang}.

\section{$Z_2$ Line-nodes and their surface states}
A $Z_2$ line-node is a line degeneracy at each point on the line there
is a linear crossing through which the parity of the Bloch states
is interchanged. Since a line-node is a time-reversal invariant object
(it includes both a ${\vec k}$ point and its time-reversal
partner $-{\vec k}$), line-nodes do not have to emerge in 
pairs. Line-nodes in our PhC are protected by the 
mirror symmetry $M_z$ and the nonsymmorphic screw
symmetries as elaborated below. The $Z_2$ topological charge of a
line-node is defined as\cite{yige} 
\be
{\cal N}_{LN} = \frac{1}{2}(m_i - m_o) . \label{nzl}
\ee 
Here $m_i$ and $m_o$ denote the eigenvalues of the mirror symmetry 
that protects the line-node for ${\vec k}$ inside and outside the
line-node, respectively. They are calculated for the band with lower
frequency that involves in the line-node band-crossing. 
${\cal N}_{LN}$ can be $\pm 1$, if the line-node is topologically nontrivial.

\begin{figure}
\begin{center}
\includegraphics[width=8.6cm]{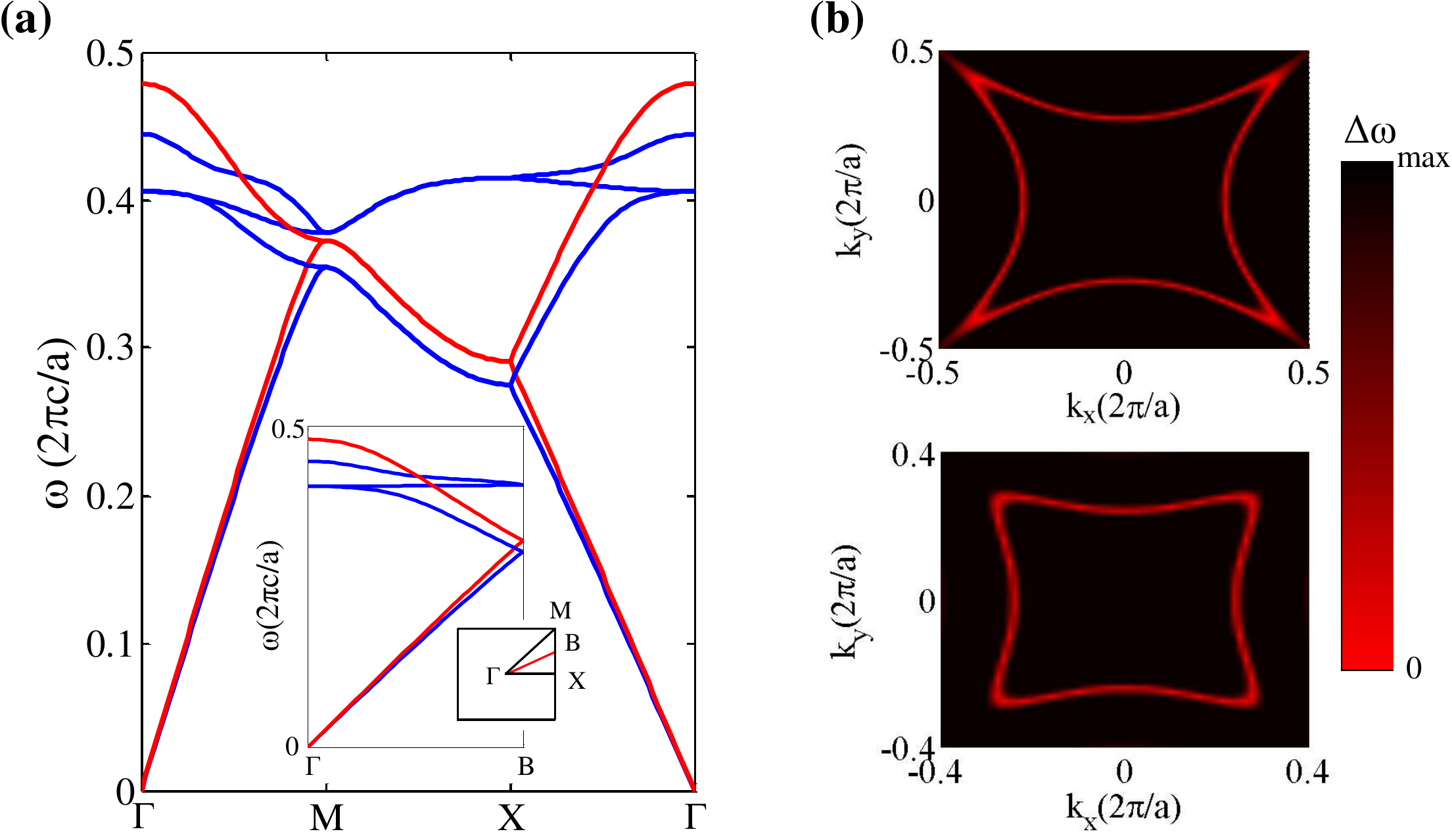}
\caption{ (Color online) (a) Photonic band structure for $k_z=0$ plane. Red curves
  label bands with $m_z=-1$, while blue curves label bands with
  $m_z=1$. Inset: band structure on $\Gamma$-B line. The B point (with
  wavevector $(0.5,0.25,0)\frac{2\pi}{a}$) represents a generic,
  non-symmetric point on the X-M line. (b) Upper: frequency
  difference $\Delta\ome$ between the fourth band and 
  the sixth band; Lower: frequency difference between the fifth band
  and the sixth band in the $k_z=0$ plane. Bands are numbered according
  their frequencies near the $\Gamma$ point in ascending orders.} 
\end{center}
\end{figure}

The line-nodes are on the $k_z=0$ plane which is invariant under $M_z$.
$M_z$ commute with both ${\Theta}_x$ and ${\Theta}_y$ when
$k_z=0$. If $\ket{m_z}$ labels a Bloch state which is an
eigenstate of the $M_z$ operator with eigenvalue $m_z$, then
\be
M_z\Theta_i \ket{m_z} = \Theta_i M_z \ket{m_z} = m_z \Theta_i \ket{m_z},
\ee
for $i=x,y$. The above equation indicates that the degenerate
partners, $\ket{m_z}$ and $\Theta_x \ket{m_z}$, at the X-M line has
the {\em same} $M_z$ eigenvalue. Due to the intrinsic property of Maxwell
equations, in the long wavelength limit $\lambda \gg a$, the photonic
bands become TM and TE modes which have opposite $M_z$
eigenvalues. These two modes become degenerate in the zero frequency
limit. When such degenerate bands with {\em opposite} $M_z$
eigenvalues evolve to some generic point on the X-M 
line and then evolve back to the $\Gamma$ point, there must be at
least one crossing of bands with opposite $M_z$ eigenvalues [see
Fig.~5(a)]. Hence the degenerate-partner switching between bands with
opposite $M_z$ eigenvalues leads to a {\em deterministic} line-node.
Here the lowest line-node emerge as the deterministic one. 
Higher line-nodes can in principle be removed by continuously tuning
the photonic bands without changing the symmetry of the PhC. The
lowest line-node is then protected 
by the mirror symmetry $M_z$ as well as the nonsymmorphic symmetry
$\Theta_x$.

Higher line-nodes can be understood through the nature of degeneracy
for higher bands at $\Gamma$ point. At $\Gamma$ point, all
nonsymmorphic operators commute with each other as ${\vec k}=0$. The
symmetry group of Bloch states then becomes equivalent to $C_{4v}
\otimes M_z$. For instance, the third and fourth bands 
are degenerate due to the $C_{4v}$ group. The two bands are of the
same $M_z$ eigenvalue ($m_z=1$) and of 
$p$-wave nature in the $x$-$y$ plane. The fifth band is also of
$m_z=1$, however, it is of $d$-wave nature in the $x$-$y$ plane and
hence non-degenerate. As a consequence, there emerge another two
line-nodes as the odd parity band crossing the other two even parity
bands [see Fig.~5(a)]. To confirm that these crossings are indeed
line-nodes, we plot the frequency difference between the fourth band
and the sixth band, as well as that between the fifth band and the
sixth band for the $k_z=0$ plane in Fig.~5(b). Here bands are numbered
according their frequencies near the $\Gamma$ point in ascending
order. From the figure, we confirmed that there are two higher
line-nodes on the $k_z=0$ plane, beside the deterministic
line-node.

The topological charge of each line-node is defined by Eq.~(\ref{nzl}). We
find that the lowest line-node is of ${\cal N}_{LN}=-1$, whereas the
second and third line-nodes have topological charge ${\cal N}_{LN}=1$.
The bulk topology dictates that there are surface states on the (001)
surface associated with these line-nodes. We remark that on (001)
surface the $Z_2$ DPs have no effect, since two DPs
with opposite $Z_2$ charge project onto the same point in surface BZ. 

\begin{figure}
\begin{center}
\includegraphics[width=8.6cm]{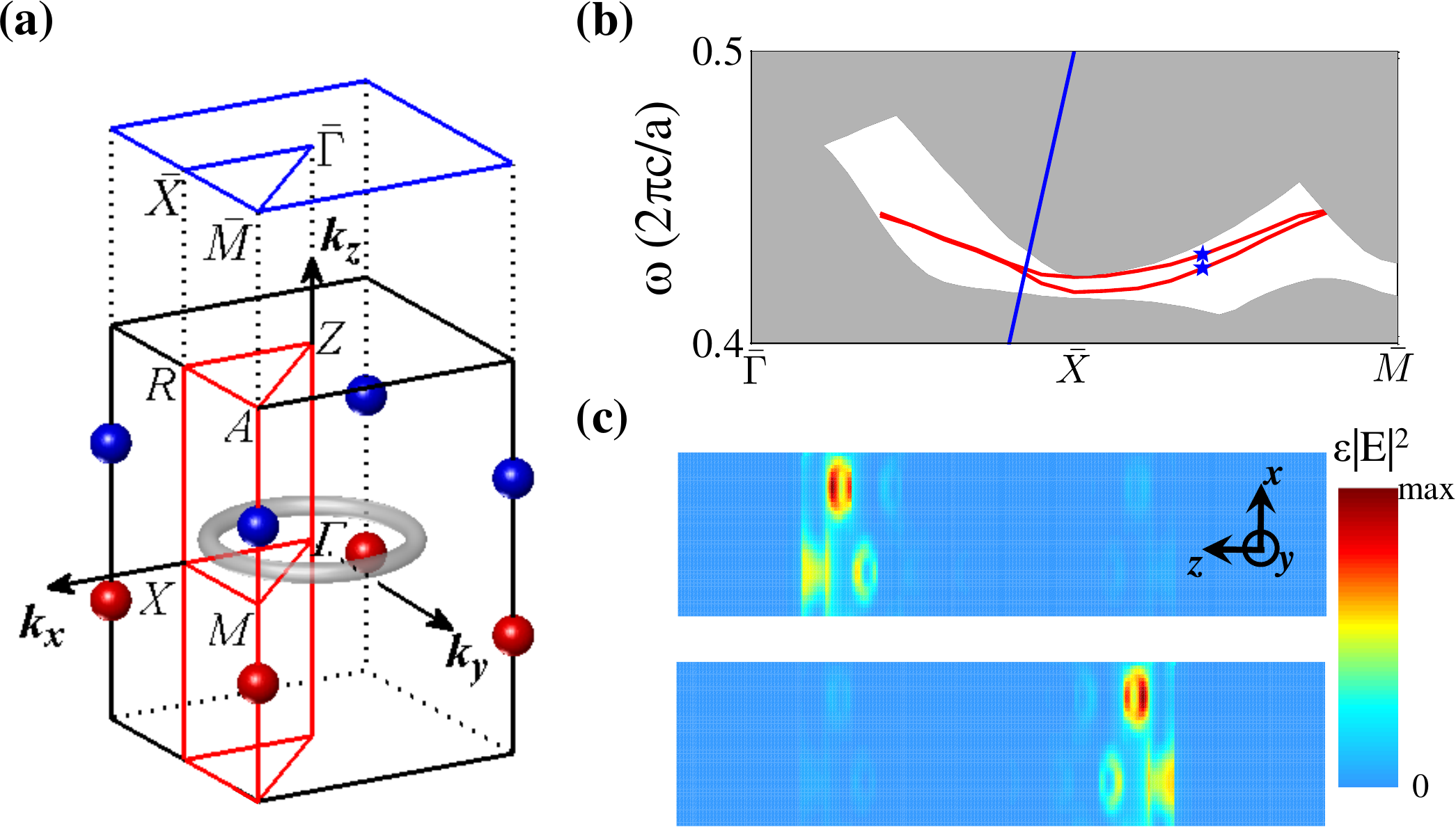}
\caption{ (Color online) (a) Line-nodes and surface BZ for
  (001) surface. (b) Spectrum of the edge states (red curves) and projected bulk bands
  (gray regions). Two nearly degenerate surface states are labeled by two blue
  stars. The blue line represents the light-line. (c) Electromagnetic wave energy profiles for the two surface
  states in the $x$-$z$ plane. The wavevectors of the two surface
  states are the same $(0.5,0.2,0)\frac{2\pi}{a}$.}
\end{center}
\end{figure}

To reveal the topological surface states on the (001) surface, we 
calculate the spectrum of the surface states using a supercell with
seven periods of our PhC unit cells along $z$ direction (details are
given in Appendix~B). From Fig.~6(b), there are two nearly degenerate
branches of surface states. Each of them is associated with one of the
interface [see Fig.~6(c)]. The small splitting between them is an
artifact of finite size calculation. We emphasize again that part of
the surface states are below the light-line and hence can exist on the
(001) PhC-air interface.

\section{Applications}
\label{sec:app}
\subsection{Ultrastrong light-matter interaction at photonic-crystal--air interface}

An important application of the $Z_2$ topological states in our
PhCs is to realize an ``open cavity'' at the interface between PhC and
air [see Fig.~7(a)]. There are topological surface states below the
light-line [see Figs.~4(b) and 6(b)]. Those surface states are
confined at the PhC-air interface without additional light-trapping
mechanism. We will show that such configuration is highly favorable for the
study of strong light-matter interaction as well as for experimental 
fabrication and measurements. First, since air has lower dielectric
constant than a dielectric mirror, the electric field in the air
region close to the interface is much enhanced. The
electric field profile in Fig.~7(a) illustrate the strong confinement
of light at the PhC-air interface. Such strong light focusing can
leads to very strong light-matter interaction which is crucial for the
study of quantum nature of electromagnetic waves.

\begin{figure}
\begin{center}
\includegraphics[width=8.6cm]{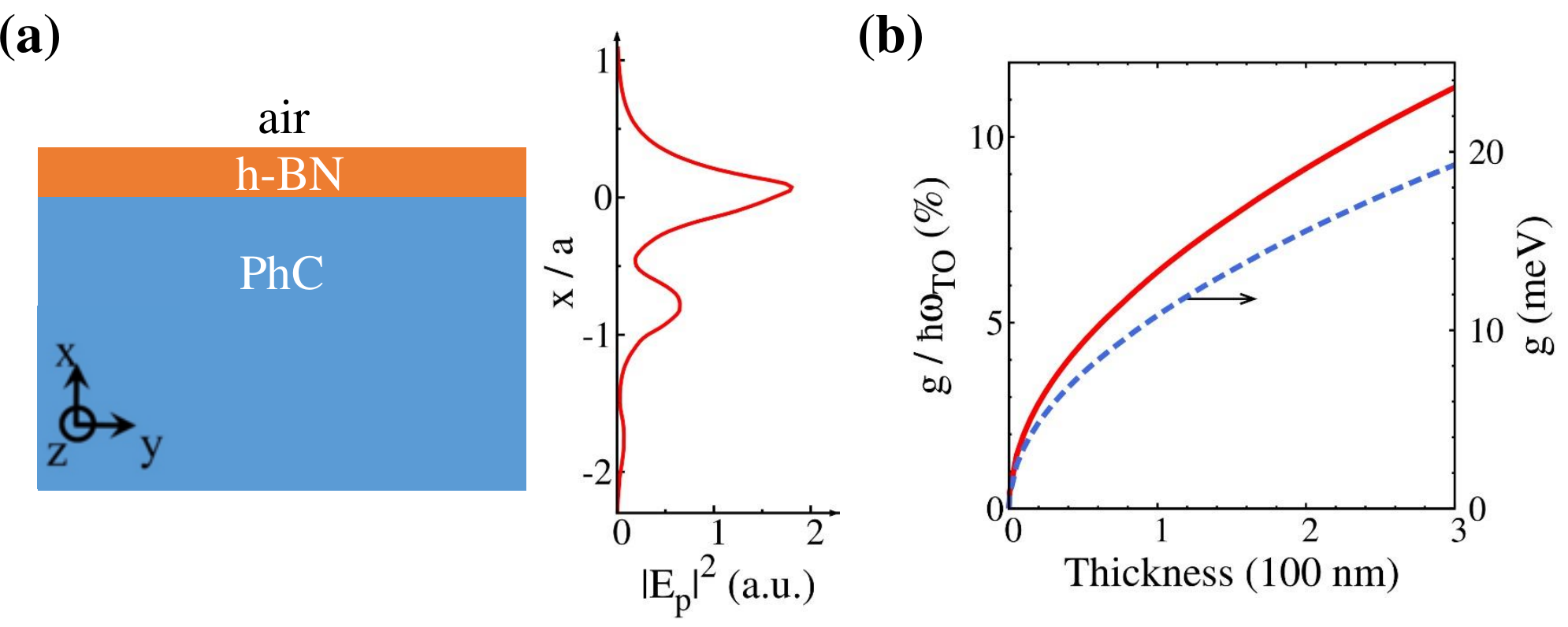}
\caption{ (Color online) (a) Open cavity at the interface between
  PhC and air. Left panel: a thin film of hexagonal boron nitride is
  placed on top of the PhC. Right panel: distribution of intensity of
  inplane electric field $a^{-2}\int dy dz |E_p({\vec r})|^2$
  over the $x$ direction. Interface between PhC and h-BN is at
  $x=0$. (b) Interaction strength between TO-phonon and photon $g$ and
  the ratio $g/(\hbar\ome_{TO})$. The interaction is calculated for
  the surface state with wavevector ${\vec
    k}=(0,0.5,-0.3)\frac{2\pi}{a}$ which is in resonance
  with TO-phonon if $a=3~\mu$m. TO-phonon has energy
  $\hbar\ome_{TO}=170$~meV in h-BN. }
\end{center}
\end{figure}

To demonstrate the merits of the surface states, we calculate the
interaction between transverse-optical(TO)-phonon in hexagonal boron-nitride (h-BN) and
cavity photon for the situation when a thin film of h-BN is placed on
top of the (100) PhC-air interface. The interacting phonon-photon system
is described by the following Hamiltonian,
\begin{subequations}
\begin{align}
& H = \sum_{\vec k}\big[ \hbar\ome_{TO}({\vec k})
b^\dagger_{\vec k} b_{{\vec
      k}} + \hbar\ome_{\vec k} a_{\vec k}^\dagger
  a_{ \vec k} \big] + H_I , \\
& H_I = \sum_{\vec k}  g_{\vec k} (b^\dagger_{\vec k} a_{\vec
  k}+a^\dagger_{\vec k}b_{\vec k}) , 
\end{align}
\end{subequations}
where $\ome_{TO}({\vec k})$ is the dispersion of the TO phonon which
is almost flat as compared with the photonic dispersion $\ome_{\vec
  k}$ of the topological surface states. $b^\dagger_{\vec k}$ creates a TO
phonon in the h-BN thin film with wavevector ${\vec k}$, whereas
$a_{\vec k}^\dagger$ creates a photon in the open cavity.
The collective coupling between phonon and photon is calculated as\cite{prx,sr}
\be
g_{\vec k} = \frac{\sqrt{ L_c  s^2
}}{2} \sqrt{\sum_l S_{u.c.}^{-1} \int_{u.c.}
  dydz |E_p(y,z,x_l)|^2}, 
\ee
where $L_c=0.47$~nm is the thickness of a single h-BN monolayer,
$\hbar\ome_{TO}=170$~meV is the TO-phonon energy,
the coupling coefficient is $s^2=3.49\times 10^{10}(2\pi)^2
c^2\hbar^2$ (in unit of Joule$^2$) as determined by experimental 
measurements\cite{geick}. Here $u.c.$ represents the ``unit-cell'' in
the $y$-$z$ plane, and $S_{u.c.}=a^2$ is the area of the unit cell.
The integral in the above equation is performed over the
strength of the electric field along the $y$-$z$ plane (i.e.,
parallel to the h-BN monolayer plane) $|E_p|^2=|E_y|^2+|E_z|^2$, since
polarization of the TO phonon is along the $y$-$z$ plane. $x_l$
is the position of the $l$-th h-BN monolayer. The electric field here
is normalized as $S_{u.c.}^{-1}\int_{u.c.}d{\vec r}
\ep(\vec{r})|{\vec E}({{\vec k}}, {\vec r})|^2 = 1$ for the surface
states. 

The interaction $g$ is calculated for ${\vec
  k}=(0,0.5,-0.3)\frac{2\pi}{a}$. The dependence of $g$ on the
thickness of the h-BN thin film is shown in Fig.~7(b). The interaction
increases approximately square root of the thickness of the thin
film, since the collective coupling is proportional to the square root
of number of h-BN monolayers. With a h-BN thin film of thickness 300~nm
(only $0.1a$), the phonon-photon coupling $g$ reaches to
19~meV. The vacuum Rabi splitting is as large as 23\% of the phonon
frequency, signifying the {\em ultrastrong} coupling
regime\cite{ciutirmp}. We have chosen the lattice constant $a=3~\mu$m
to ensure that the TO-phonon is in resonance with the surface photon.

In quantum optics, when the ratio $g/\hbar\ome> 0.1$, the
system enters into the ultrastrong coupling
regime\cite{ghz,thz1,thz2,mir}. There are several
nontrivial properties of an ultrastrong coupling system. For instance, 
the ground state of an ultrastrong coupling system has a finite 
population of bound photons which enables quantum vacuum
emission\cite{qve}. The ultrastrong coupling system can also be
exploited to study strong single photon nonlinearity and quantum phase
transition\cite{block}. Nontrivial Berry phases and strong interaction
are two key elements of interacting symmetry-protected 
topological states, such as fractional quantum Hall
states\cite{fqhem,wenbook} and bosonic topological
insulators\cite{wenscience}. However, it is rather 
challenging to realize such nontrivial states in known physical
systems. Our PhC open-cavity with phonon-polaritons in the ultrastrong
coupling regime in mid infrared frequency might open a route to such
states since both strong interaction and nontrivial Berry phases can
exist in our system. Future studies should address the relevant
parameter regimes for quantum simulation of strongly interacting
topological phonon-polariton gas.

Furthermore, fabrication of such cavity can be much easier than the
sandwich-shaped cavity as studied in the literature for both
Fabry-P\'erot cavities\cite{ciutirmp,sr} and PhC
cavities\cite{prx}. The PhC studied here can be fabricated using 
mature commercial technology such as
$\copyright$NanoScribe. After that the h-BN thin film can be 
fabricated separately and transferred to the surface. Moreover, it is
easier to study the physical properties of the ultrastrong coupling
phonon-photon system on the PhC-air interface. Instruments such as
near field scanning can be useful for Hanbury Brown-Twiss
interferometry and high-order photon correlation
measurements. 

\subsection{Anomalous refraction and valley physics of type-II $Z_2$
  Dirac Points}
In our PhC, the DPs are on the M-A line with
$k_x=k_y=\frac{\pi}{a}$. Through a surface 2D grating with periods
along both $x$ and $y$ directions being $2a$, light from air can directly couple
to the DPs. It is equivalent to regard the effect of the
surface grating as duplicating the light cone of air at ${\vec
  k}=(0,0,0)$ to $(\frac{\pi}{a}, \frac{\pi}{a}, 0)$. The refraction
of this duplicated light cone through the DPs is the focus of
study in this section [see Fig.~8(a)]. We will show that type-II $Z_2$
Dirac cones (i.e., highly tilted Dirac cones) lead to anomalous refraction
and valley physics.

In our PhC the main deformation of Dirac cones comes from a finite
velocity along the $z$ direction. Taking into account of such
deformation, photonic dispersion around a DP is written as 
\be
\ome = \ome_0 + v ( \eta q_z \pm q) .
\ee
Here $\ome_0$ is the frequency at the DP, $v$ is the group
velocity of the DP. The dimensionless parameter $\eta$
measures the degree of tilt of the Dirac cone along $q_z$
direction. Here ${\vec q}=(q_x,q_y,q_z)$ represents the deviation of
the wavevector away from the DP at
$(\frac{\pi}{a},\frac{\pi}{a}, k_{DP})$ and
$q=\sqrt{q_x^2+q_y^2+q_z^2}$.

\begin{figure}
\begin{center}
\includegraphics[width=8.6cm]{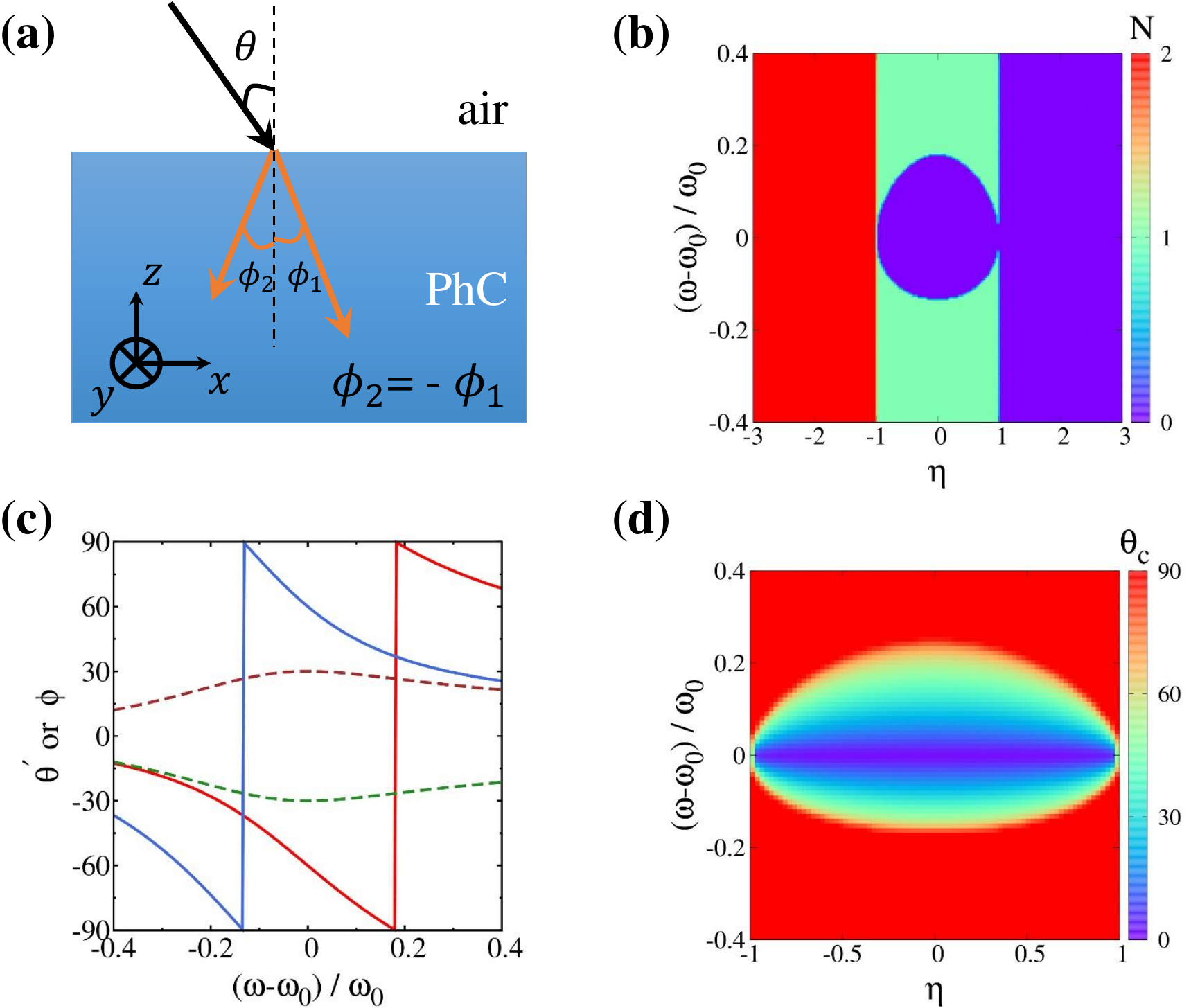}
\caption{ (Color online) (a) Anomalous refraction of a type-II
  DP. A light beam from air (black arrow) is shed on the PhC,
  which is refracted into two beams in the PhC with different group velocities
  (orange arrows). The two refraction angles are always {\em opposite} to
  each other, $\phi_2=-\phi_1$. (b) Number of refraction beams $N$ for
  various frequency $\ome$ and $\eta$ when the incident angle is
  $\theta=50^\circ$. There are two refraction beams for $\eta<-1$,
  whereas no refraction beam for $\eta>1$. (c) 
  The angles for the wavevectors $\theta^\prime$ and group velocities
  $\phi$ of the two refraction beams in PhC as functions of frequency
  $\ome$ for incident angle $\theta=50^\circ$ and parameter $\eta=-2$. The red
  (blue) curve represents $\theta^\prime_1$ ($\theta^\prime_2$), the
  brown (green) curve represents $\phi_1$ ($\phi_2$). (d) Critical
  angle $\theta_c$ for type-I Dirac cone vs. frequency and $\eta$. 
  Parameters: $\ome_0=0.5\frac{2\pi c}{a}$ and $v=0.2c$.}
\end{center}
\end{figure}

The refraction beams are determined by the frequency $\ome$ and
incident angle $\theta$ of the light. Since the dispersion is
isotropic in the $q_x$-$q_y$ plane, we can simplify the problem by
setting $q_y=0$. The conservation of frequency and $q_x$ determines
the angle $\theta^\prime$ for the refraction beam,
\begin{align}
& \ome - \ome_0 = v (\eta q_z \pm \sqrt{q_z^2 + q_x^2}), \nn\\
& q_x = \frac{\ome\sin\theta}{c}, \quad \theta^\prime \equiv -\arctan\frac{q_x}{q_z}
, \label{refrac}
\end{align}
where the incident angle $\theta$ is positive in our discussions.
The above equations may have multiple solutions or no solution. The
latter signifies no refraction through the DP, or that such a
DP cannot be excited by light from air. For the case with
multiple solutions, only the solutions with 
group velocity along $z$ direction $v_z<0$ delivers the refraction
beam, where 
\be
v_z = v \left( \eta \pm \frac{q_z}{\sqrt{q_x^2 + q_z^2}} \right) .
\ee
From the above equation we find that for $\eta>1$, refraction
through the DP is forbidden, since both branches have
positive velocity. On the other hand, for $\eta<-1$,
there are two solutions for Eq.~(\ref{refrac}) [see Fig.~8(b) for the
number of refraction beams for various $\eta$ and frequency when the
incident angle is $\theta=50^\circ$]. Indeed we observe
birefringence as shown in Fig.~8(a). The two angles $\theta_1^\prime$
and $\theta_2^\prime$ as a function of frequency is shown in
Fig.~8(c) for incident angle $\theta=50^\circ$, $\ome_0=0.5\frac{2\pi
  c}{a}$, and $v=0.2c$. We notice that $q_z$ can switch sign for both
refraction beams (as indicated by abrupt change in
$\theta^\prime$). An important quantity of the refraction beams is the 
group velocity, consisting of both $z$ and $x$ components,
\be
v_x = \pm v \frac{q_x}{\sqrt{q_x^2 + q_z^2}}, \quad \phi \equiv
-\arctan\frac{v_x}{v_z} .
\ee
From Fig.~8(c) we observe a rather simple relation between the two
refraction beams,
\be
\phi_1=-\phi_2.
\ee
The above relation indicates that one of the refraction beam has
negative refraction. The above relation emerges because for $\eta<-1$
there is a solution for each branch ($\pm$) with
\be
q_{z,\pm} = \frac{\eta(\ome-\ome_0)\pm
  \sqrt{(\ome-\ome_0)^2+v^2(\eta^2-1)q_x^2}}{v(\eta^2-1)} .
\ee
We emphasize that the magnitude of the two group velocity is
different, unless $\ome=\ome_0$. 
Because DPs appear in pairs with opposite
$\eta$ due to time-reversal symmetry. For type-II DPs, one of
the Dirac cone cannot be excited, whereas the other one exhibits
birefringence. This interesting phenomenon provides a route to valley
selective physics for photonic DPs. We emphasize that when
light is incident from the bottom surface (instead of the top surface)
of the PhC, opposite valley contrast is realized, since in this
geometry the refraction beams must have $v_z>0$. 

In contrast, for type-I DPs, excitation of the Dirac cone is
possible if the incident angle is smaller than a critical angle
$\theta_c$. If $\theta<\theta_c$ there is only one refraction beam. We
plot $\theta_c$ for various frequency and $\eta$ in Fig.~8(d). From
the figure we find that the critical angle $\theta_c$ is the
same for both negative and positive $\eta$ with the same $|\eta|$ at
the same frequency. Therefore, type-I DPs have no valley contrasting in
refraction.

\section{Conclusions and discussions}

We propose an effective method for simulation of 3D $Z_2$ topological
nodes in all-dielectric PhCs with space-time reversal symmetry using
nonsymmorphic screw symmetries. In a concrete example we show that for
a tetragonal lattice with space group of P4$_2$/mcm, the screw
symmetries lead to double degeneracy in high symmetry planes in the
BZ. In this way, the screw symmetries play a similar role as Kramers
degeneracy for spin-1/2 electronic $Z_2$ topological states. 

Using this method, we find a number of $Z_2$ DPs on the
M-A line (i.e., the common line of the doubly degeneracy planes).
We also find $Z_2$ line-nodes around the $\Gamma$ point on the
$k_z=0$ plane for the first few photonic bands. The lowest line-node
emerges in a deterministic manner because of a {\em degenerate-partner
switching} mechanism: the degenerate-partners have the opposite
parities at the $\Gamma$ point due to TE-TM degeneracy at zero
frequency but have the same parity on the boundaries of the BZ on
the $k_z=0$ plane due to nonsymmorphic screw symmetry. Hence there
must be a line crossing of bands during the evolution between the
$\Gamma$ point and BZ boundaries.

As a consequence of $Z_2$ DPs, a pair of Fermi arcs with opposite
chirality are found on the (100) surface of the PhC. The spectrum of
such double Fermi arcs is mirror symmetric with respect to the 
$k_y=\pi/a$ line. The surface states are protected by the screw
symmetry $S_y$ which guarantees double degeneracy for
$k_y=\pi/a$ in the surface BZ. Remarkably, parts of 
the surface states are below the light-line which hence allows
realization of an ``open cavity'' on the PhC-air interface. We find that
light is strongly confined around the interface due to the topological
surface states, which allows {\em ultrastrong coupling} between TO-phonon in
h-BN thin film and cavity photons. Such realization is feasible within
current technology and has several fabrication and measurement
advantages. Our ultrastrong coupling system provides an emergent
platform for the study of {\em extreme limits} in quantum optics and quantum
phase transitions. With both strong interaction
and nontrivial Berry phases, the system may host nontrivial states
of matter such as strongly interacting bosonic topological
states\cite{wenscience,fqhem}. Furthermore, TO-phonons in h-BN fall
into the {\em mid infrared frequencies}, $170$~meV, which is also important in
manipulation of near field thermal emission\cite{thermal,thermal2}.

We also find that type-II $Z_2$ DPs have anomalous valley selective
refraction: birefringence for one valley, while no refraction for the
other. For the valley with birefringence, the two refraction
beams have {\em opposite} refraction angles, one positive and the
other negative. These findings open a window for the study of 
valley physics\cite{valley1,valley} in Dirac and Weyl photonic
systems.

\section*{Acknowledgments.} 
Researches in Soochow University are jointly supported by the faculty
start-up funding and the National Science Foundation of China (Grant
nos: 61322504 and 11574226). J.H.J thanks Sajeev John for many helpful
and illuminating discussions. Y.C and H.Y.K are supported by NSERC of
Canada and Center for Quantum Materials at the University of Toronto.

\appendix

\section{${\vec k}\cdot{\vec P}$ analysis of DPs}

We use the ${\vec k}\cdot{\vec P}$ theory for photonic bands to study
the low-energy Hamiltonian of the DPs. The photonic bands are
solutions of the following eigenvalue equations
\be
\grad\times \frac{1}{\vep({\vec r})} \grad\times {\vec
  h}_{n,\vec{k}}({\vec r}) = \frac{\ome_{n,\vec{k}}^2}{c^2} {\vec
  h}_{n,\vec{k}}({\vec r}) ,
\ee
where $n$ is the band index and ${\vec h}_{n,\vec{k}}({\vec r})$ is
the Bloch function for the magnetic field of the electromagnetic wave.
It is normalized as $\int_{u.c.} d{\vec r} {\vec
  h}_{n^\prime,\vec{k}}^\ast({\vec r}){\vec h}_{n,\vec{k}}({\vec
  r})=\delta_{nn^\prime}$ with $u.c.$ denoting the unit cell (i.e.,
integration within a unit cell). The Hermitian operator $\grad\times
\frac{1}{\vep({\vec r})} \grad\times$ is then regarded as the
``photonic Hamiltonian''\cite{our}.

The ${\vec k}\cdot{\vec P}$ theory is constructed by expanding the
Bloch functions of photonic bands near the DP with the Bloch
functions at the DP. Direct calculation yields,
the following ${\vec k}\cdot{\vec P}$ Hamiltonian,
\begin{widetext}
\be
{\cal H}_{nn^\prime} ({\vec k}) = \frac{\ome_{n,0}^2}{c^2}\delta_{nn^\prime} +
{\vec q}\cdot {\vec P}_{nn^\prime}  - \int_{u.c.} 
\frac{d{\vec r}}{\vep({\vec r})} \vec{h}_{n,0}^\ast({\vec
  r})\cdot[{\vec q}\times({\vec q}\times \vec{h}_{n^\prime,0}({\vec r}))] ,
\ee
where $\ome_{n,0}$ is the eigen-frequency of the $n^{th}$ band at the
DP. The matrix element of ${\vec P}$ is given by 
\be
{\vec P}_{nn^\prime} = \int_{u.c.}\frac{d\vec{r}}{\vep({\vec r})} [\vec{h}_{n^\prime,0}({\vec
  r})\times(i\grad\times \vec{h}_{n,0}^\ast({\vec r}))+(i\grad\times
\vec{h}_{n^\prime,0}({\vec r}))\times \vec{H}_{n,0}^\ast({\vec
  r})]  .
\ee
A crucial fact is that the matrix element ${\vec P}_{nn^\prime}$ is
nonzero only when the $n$ and $n^\prime$ bands are of different
parity. In our system the mirror planes are $y=x$, $y=-x$, and
$z=0.25$. Therefore, we shall calculate the two momentum matrix
elements, $\hat{P}_{1}=(\hat{P}_x+\hat{P}_y)/\sqrt{2}$ and $\hat{P}_{2}=(-\hat{P}_x+\hat{P}_y)/\sqrt{2}$.

We calculate the ${\vec P}$ matrix element for the second DP
at $k_z>0$. The matrix form (band index goes from 3 to 6) gives (in
arbitrary units), 
\bea
&& \hat{P}_{1} = \left( \begin{array}{cccccccccccc}
    0 & 0 & 0 & 0  \\
    0 & 0 & -0.35 -0.26 i &  -0.40 + 1.61 i\\
    0 & -0.35 + 0.26 i & 0 & 0 \\
    0 & -0.40 - 1.61i & 0 & 0 \\
    \end{array}\right) , \nn \\
&& \hspace{-2cm} \hat{P}_{2} = \left( \begin{array}{cccccccccccc}
    0 & 0 & 1.65 + 0.04i & 0.16 + 0.4 i \\
    0 & 0 & 0 & 0 \\
    1.65-0.04i & 0 & 0 & 0 \\
    0.16 - 0.4i & 0 & 0 & 0 \\
    \end{array}\right) , \quad \quad\quad
\hat{P}_{z} = \left( \begin{array}{cccccccccccc}
    -0.74 & 0 & 0 & 0 \\
    0 & -0.74 & 0 & 0 \\
    0 & 0 & 1.5 & 0 \\
    0 & 0 & 0 & 1.5  \\
    \end{array}\right) . 
\eea
\end{widetext}
The ${\vec k}\cdot {\vec P}$ Hamiltonian in the ${\vec k}$ linear
order is then
\be
\hat{{\cal H}}- \frac{\ome_0^2}{c^2}\hat{1} \sim q_1 \hat{P}_1 + q_2 \hat{P}_2 + q_z \hat{P}_z ,
\ee
where $q_1=(q_x+q_y)/\sqrt{2}$ and
$q_{2}=(-q_x+q_y)/\sqrt{2}$, $q_x=k_x-\pi/a$, $q_y=k_y-\pi/a$, and
$q_z=k_z-k_{DP}$. The above Hamiltonian faithfully restore the
symmetry and properties of the DP: (i) There are two doublets
for each ${\vec q}$ [up to numerical error ${\cal O}(0.01)$]. (ii)
Interaction is finite only between states of opposite
parity. That is, the first two states have opposite parity with the
other two states.

\begin{figure}
\begin{center}
\includegraphics[width=8.6cm]{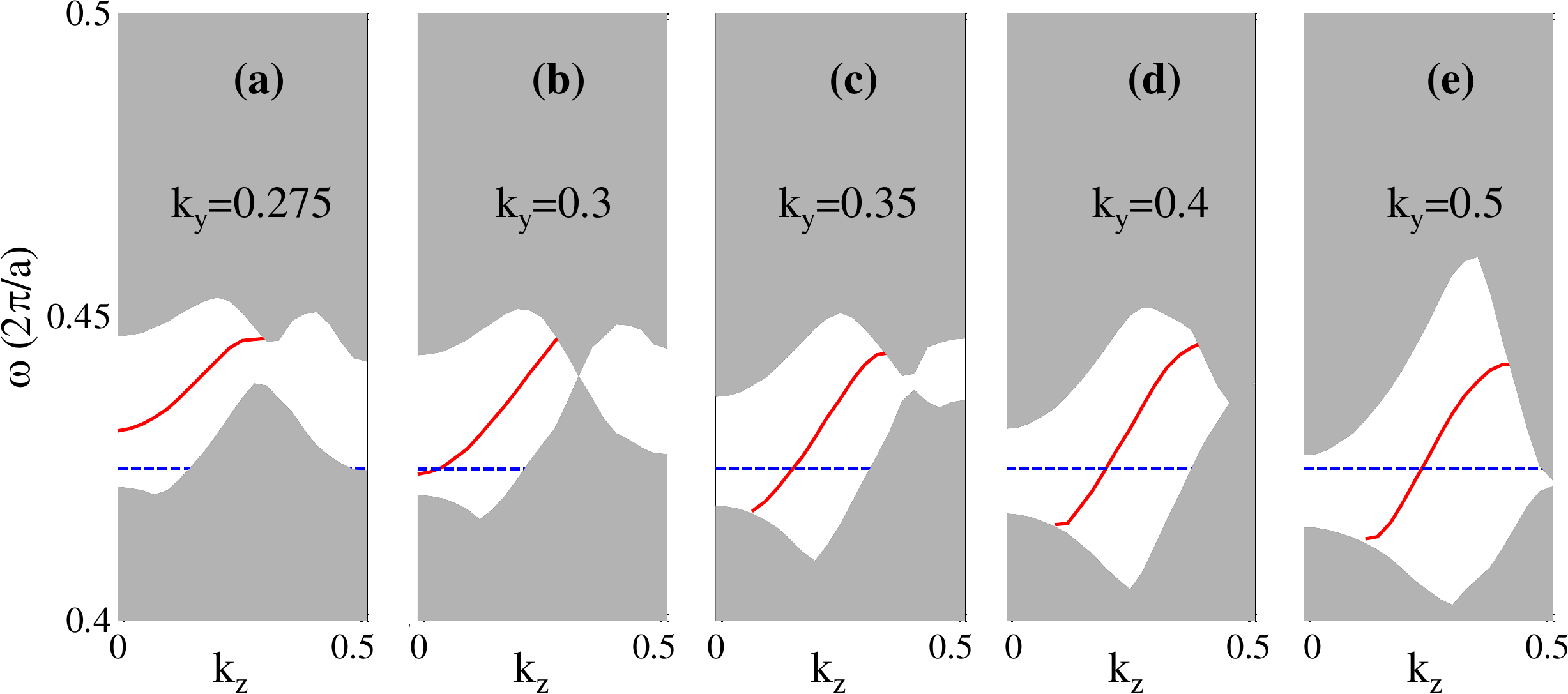}
\caption{ (Color online) (a)-(e) Projected bulk bands (gray regions) and
  surface spectrum (red curves) vs. $k_z$ for different $k_y$. All
  $k_z$ and $k_y$ are in units of $2\pi/a$. Frequencies are in unit of
  $\frac{2\pi c}{a}$. The blue line indicate the
  frequency $\ome=0.425\frac{2\pi c}{a}$.}
\end{center}
\end{figure}

\begin{figure}
\begin{center}
\includegraphics[width=8.6cm]{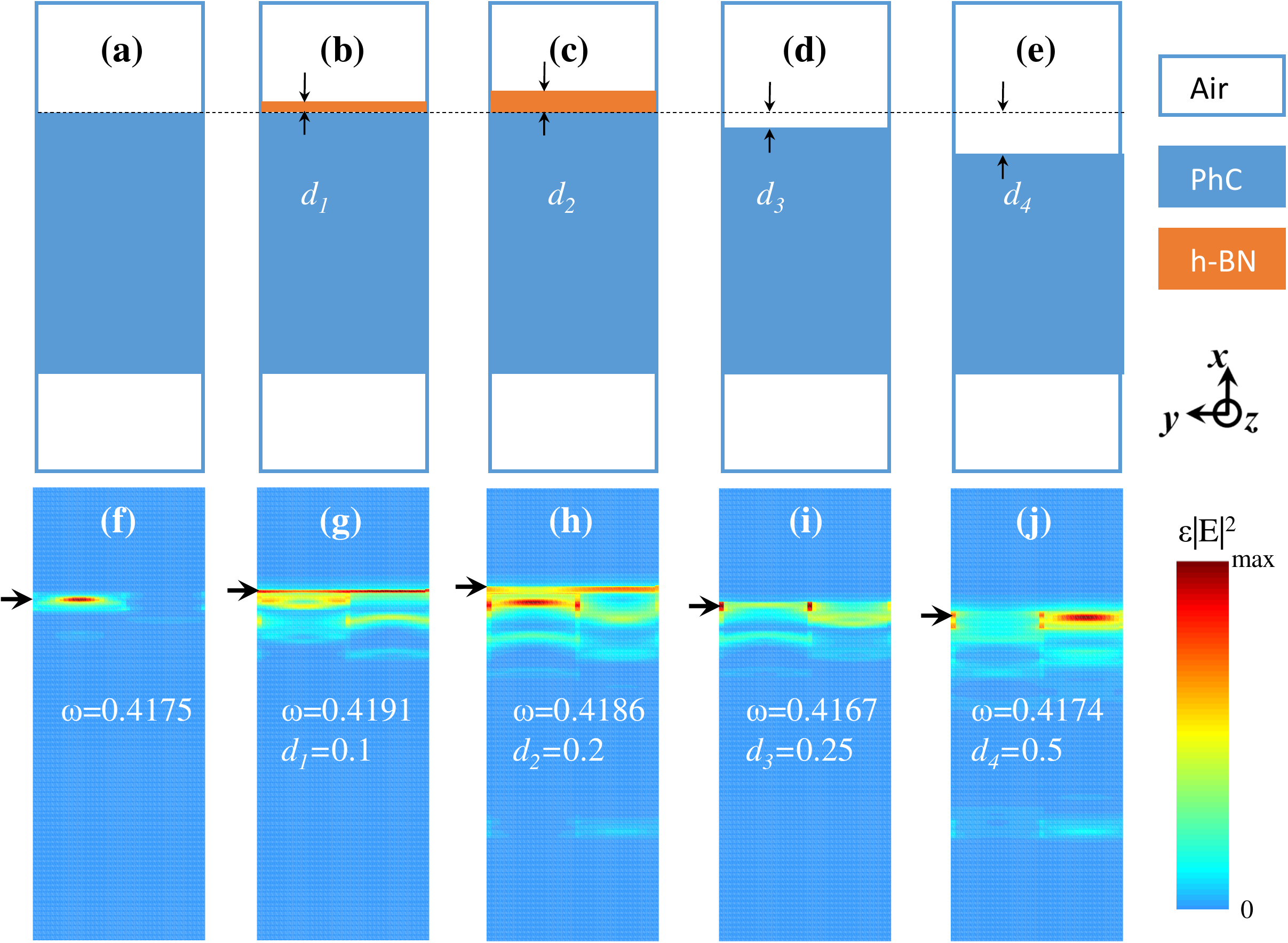}
\caption{ (Color online) (a)-(e): Schematic of various modification of
  the supercell used in calculation of surface states. (a) Seven
  periods of PhC stacking along the $x$ direction and $3a$ thick air
  layers above and below the PhC. (b) A layer of h-BN (thickness
  $0.1a$) is placed on top of the PhC. (c) A layer of h-BN of
  thickness $0.2a$ is placed on the top surface. (d) The top surface layer
  of PhC is cut by $0.25a$. (e) The top surface layer is cut by $0.5a$.
  (f)-(j): Frequency and energy profile $\vep|\vec{E}|^2$ (integrated over
  $z$ direction) in the $x$-$y$ plane of the surface state with ${\vec k}=(0,0.5,-0.3)\frac{2\pi}{a}$ for various
  situations. The arrows
  indicate the position of the top interface.}
\end{center}
\end{figure}

\section{Calculation of Fermi arcs associated with $Z_2$ DPs}
To calculate the Fermi arcs on the (100) surface, we set a supercell
with seven periods of PhC stacked along $x$ direction. At the two
sides of the PhC is an effective medium with $\vep=0.2$. This gives an
effective refractive index $n=0.45$ which promotes the light-line to
higher frequency. This technique is to ensure all the surface states
of interest is confined to the surface. Without this treatment a
portion of surface states is above light-line and are not confined to
the surface any more. Similar skills have been employed to calculate the
surface states on the (001) surface which are associated with $Z_2$
line-nodes.

We emphasize that the $S_y$ symmetry is kept for our supercell
computation by proper truncation of the structure [Similarly, the
$M_z$ symmetry is kept for the supercell calculation of (001)
surface states]. The $S_y$ symmetry is crucial in keeping
the double degeneracy on the $k_y=\pi/a$ line in the surface BZ.
The Fermi arcs are obtained by scanning the surface BZ. Specifically,
for a series of $k_y$ points, we calculate the projected bulk bands,
as well as the spectrum of the surface states. The latter is obtained
by comparing the spectrum of the supercell and the projected bulk
bands. The states lie in the projected band gap are identified as the
surface states. We show in Fig.~9 the spectrum of the projected bulk
bands (gray) and surface states (red). The Fermi arc is obtained via
finding the ${\vec k}$ points in the surface BZ with the given
frequency (i.e., the intersection between the red curve and the blue
line). The spectrum is symmetric for positive and negative $k_y$ (or
$k_z$), hence we only need to compute a quarter of the surface BZ.
We remark that from Figs.~9(a) and 9(b), one can infer that the Fermi
arcs at $k_z>0$ and $k_z<0$ are continuously connected. In 
contrast, since the intersection between the blue line and surface
state spectrum is protected by topology at $k_y=\pi/a$ [the edge
spectrum traversing the ``valence'' and ``conduction'' bands as shown
in Fig.~9(e)], the intersection of Fermi arcs with $k_y=\pi/a$ line is
protected. Hence, the $k_y=\pi/a$ line separates the two Fermi arcs,
which is consistent with bulk-edge correspondence principle.

\section{Robustness of topological states on (100) surface}
To demonstrate the robustness of the topological surface states for
(100) surface associated $Z_2$ DPs, we calculate the
frequency and field profiles for various modification of the
supercell. The supercell, as demonstrated in Fig.~10(a) consists of
seven periods of our PhC and air regions of thickness $3a$ above and
below the PhC. The structure is staked along $x$ direction. There are
two PhC-air interfaces, we only focus on one of them (the top surface
shown in Fig.~10). Beside such
a supercell, we also calculate the following four situations (all for
the wavevector ${\vec k}=(0,0.5,-0.3)\frac{2\pi}{a}$ which is below
the light-line): (1) When a h-BN layer of thickness $d_1=0.1a$ is
placed on top of the PhC. The dielectric constant of the h-BN is taken
as $\vep_{BN}=8$ as from experiments\cite{geick}. (2) When the
thickness of the h-BN layer is $d_2=0.2a$. (3) When the surface layer
is cut by $d_3=0.25a$. (4) When the surface layer is cut by
$d_4=0.5a$. The geometries for those modifications are illustrated in
Figs.~10(b)-(e). For all the situations, the bottom surface is left
unchanged. The frequency of the surface state as well as its
energy distribution for various situations are shown in
Figs.~10(f)-(j). It is seen that the frequency of the surface state
changes negligibly for various situations, although the energy
distribution has been modified. These results demonstrate the
robustness of the topological surface states.

{}

\end{document}